\documentstyle[aps,prl,epsfig,amssymb]{revtex}

\def\B.#1{{\bbox{#1}}}
\def\x{{\mbox{\boldmath$x$}}}
\def\u{{\mbox{\boldmath$u$}}}
\def\r{{\mbox{\boldmath$r$}}}

\def\unitr{{\mbox{\boldmath$\hat r$}}}

\def\p{{\mbox{\boldmath$\phi$}}}

\def\n{{\mbox{\boldmath$n$}}}
\def\m{{\mbox{\boldmath$m$}}}
\def\U{{\mbox{\boldmath$U$}}}
\def\B{{\mbox{\boldmath$B$}}}
\def\W{{\mbox{\boldmath$W$}}}
\def\Z{{\mbox{\boldmath$Z$}}}

\def\unitr{{\mbox{\boldmath$\hat r$}}}

\def\la{{\langle}}
\def\ra{{\rangle}}

\newcommand{\be}{\begin{equation}}
\newcommand{\ee}{\end{equation}}
\newcommand{\bea}{\begin{eqnarray}}
\newcommand{\eea}{\end{eqnarray}}

\begin{document}
\title{Universality in passively advected hydrodynamic fields: the case of a
passive vector with pressure.}
\author{R. Benzi$^1$, L. Biferale$^1$ and  F. Toschi$^{1,2}$}
\address{$^1$Dipartimento di Fisica and INFM, 
Universita' di Roma "Tor Vergata",
Via della Ricerca Scientifica 1, I-00133 Roma, Italy}
\address{$^2$  University of Twente, Department of Applied Physics, Enschede, The Netherlands}

\maketitle
\begin{abstract}
 Universality of statistical properties 
of passive quantities advected by turbulent velocity fields at changing
the passive forcing mechanism is  discussed. In particular, 
we concentrate on the statistical properties
 of an  hydrodynamic system with pressure.
 We present  theoretical arguments and
preliminary numerical results which show that the fluxes
of passive vector field and of the velocity field have the same
scaling behavior. By exploiting such a property, we propose a 
way to compute the anomalous exponents of  three dimensional 
turbulent velocity fields. Our findings are in agreement within $5\%$
with experimental values of the anomalous exponents.

\vskip0.2cm
\end{abstract}

\vskip0.3cm
\vskip0.2cm

\section{Introduction}
Recent theoretical and numerical studies  of
passive quantities linearly advected by either stochastic 
\cite{k94,gk95,v96,abp00,pa00} 
or true turbulent Navier-Stokes velocity fields \cite{cv00}
  have focused on the 
important problem of the ``zero-mode'' dominance of the statistical 
properties of the passive field in the inertial range.\\
The field may be an advected scalar
$\theta(\x,t)$, with the equation of motion
\begin{equation}
\frac{\partial \theta}{\partial t} +\u \cdot {\bf \nabla} \theta=\kappa
\nabla^2\theta + f_{\theta} , \label{pas}
\end{equation}
or a vector, like a magnetic field $\B(\x,t)$ satisfying \cite{Zeldovich}
\begin{equation}
\frac{\partial \B}{\partial t} +\u \cdot {\bf \nabla} \B=\B\cdot
{\bf \nabla} \u + \kappa \nabla^2\B\  + {\bf f}_{B}. \label{magnetic}
\end{equation}
Zero-mode dominance is of crucial importance because naturally explains
the deviations from dimensional predictions for scaling properties and
some  universality with respect to the forcing mechanism, i.e. some 
independence of scaling properties from the large scale input
of passive field. 
 
In particular, for the special class of Kraichnan-like problems for scalar
and vectors with  Gaussian and  white-in-time velocity field and 
external forcing \cite{k94,gk95,v96,abp00}, one can prove that the equal time
(passive) correlation functions are dominated, in the inertial range,
by the zero modes of the linear  operator describing the advection
by the Gaussian velocity field.
Anomalous scaling comes from the 
  non-trivial scaling properties of the null-space
of the inertial operator. Universality comes
from the fact that {\it only} prefactor of the zero modes feel  the large
scale boundary conditions, while the power law behavior is fixed
by the inertial operator. As a consequence of zero-mode dominance 
in the inertial range we also have that all forcing-dependent
contributions are sub-dominant and therefore some degree
of universality with respect to the forcing mechanism. 
  
Recently, numerical evidences
that zero modes show up  also in true passive models, advected
by realistic non-Gaussian and non white-in-time velocity fields, have
been presented for both passive scalars in the inverse cascade regime
of a two dimensional flow \cite{cv00} and shell models
for passive scalars \cite{abp01}.  Such  findings give strong support
to the idea that statistics of quantities linearly advected by true
Navier-Stokes fields may show some universality properties with 
respect to the forcing mechanism too. In particular, the problem of 
a passive with a mean shear studied in \cite{cv00}
  shows that  zero-modes
remain  dominant with respect to the forcing contributions also in presence 
of a correlation between forcing and velocity advecting field, i.e. some
degree of universality still holds. 
In this paper we want to investigate how far one can push
the idea of universality with respect to the forcing mechanism 
for the particular case of the advection of a passive field with pressure
\cite{pa00} advected by the true Navier-Stokes turbulent velocity field.
Such a model was initially proposed in \cite{pa00} in order
to understand the importance of non-local contributions, induced
by the pressure term, in the zero-modes structure of the 
advecting operator. Here, we want to show and study  another important
striking feature  of the model, i.e. its property to describe either a
 {\it linear} or {\it non-linear}
evolution depending on the correlation between the passive forcing and
the advecting velocity fields. 
Moreover, when the passive evolution is indeed non-linear, the passive
field and the velocity fields coincide, i.e. the transported
field and the advecting field are looked in both their spatial and
temporal evolution. In other words, we have a linear model
which may become non-linear by changing the correlation properties
between the external forcing and the advecting velocity field.  

 In this paper we address two problems. First,
 we want to understand the degree of universality in the scaling
 properties   of the {\it linear}
problem  at changing the external forcing.\\   
Second, by using the similarity between the non-linear Navier-Stokes
eqs and the linear advecting-diffusive eqs for the passive field (see below) 
 we propose  a way to compute the anomalous exponents of
homogeneous and isotropic turbulence
 which takes advantage of recent results in the 
theory of passive scalar advected by Gaussian, white in time
velocity fields \cite{k94,gk95}. \\
The paper is organized as follows. In section II we present
the equations for the passive field with pressure. in section III
we present some numerical experiments made on the equivalent
shell model for the passive field and we discuss to which extent
we may expect strong universality with respect to the forcing
mechanism in this model. In particular we discuss why we expect, and
verify numerically in the shell model case, that passive
fluxes posses a much better degree of universality than passive
structure functions. In section IV we used the, supposed,  universality
of passive flux in order to derive a functional constraint for the
velocity statistics. Conclusions follow in section V. 

\section{The model}
Let us consider the incompressible Navier-Stokes (NS) equations:
\be
\label{NS}
\partial_t U_i + U_j \partial_j U_i = - \partial_i p + 
\nu \Delta U_i + f_i
\ee
where  notation follows the usual meaning and the 
forcing term $f_i$ is supposed
to produce a stationary homogeneous and isotropic turbulence.
We next consider a vector field $\W$, divergence-less 
(i.e. $\partial_i W_i = 0$) and
 satisfying a transport-like  equation:
\be
\label{NSW}
\partial_t W_i + U_j \partial_j W_i = - \partial_i p_w + 
\nu \Delta W_i + g_i
\ee
where the "pressure" term $p_w$ is computed via the Poisson equation:
\be
\partial_i U_j \partial_j W_i = - \Delta p_w
\ee
Our aim is to discuss the statistical properties of $\W$ and 
their relation
with the scaling properties of $\U$.\\

Let us  introduce the anomalous exponents 
for $\U$ and $\W$. We shall
denote by $\zeta(p)$ the scaling exponents of
 the longitudinal 
structure function: $$S^p(r) = \la (\delta_r U)^p\ra \sim r^{\zeta(p)}$$
where 
$\delta_r U = (\U ( \x+ \r)- 
\U(\x)) \cdot  \unitr$ and
$\la \cdot \ra$ stands for ensemble averaging. 
In the  same way, we introduce $\sigma(p)$ as the scaling exponents 
of longitudinal linear-field structure functions:
 $$T^p(r) = \la (\delta_r W)^p\ra \sim r^{\sigma(p)}$$
where 
$\delta_r W = (\W ( \x+ \r)- 
\W(\x)) \cdot  \unitr$.
Finally,we consider the scaling exponents
 $s(p)$ for the $\W$ {\it flux} defined as:
\be
\la (\delta_r U)^p (\delta_r W)^{2p} \ra  \sim r^{s(p)}
\ee
Let us remark that equation (\ref{NSW}) implies $s(1) = 0$ which follows from
the analogous of the "4/5" Kolmogorov equation for $\W$ \cite{frisch},
 namely
\be
\la \delta_rU (\delta_rW)^2 \ra \sim N r
\ee
where $N$ is the mean rate of $W$ dissipation. 
Our analysis is aimed at understanding
 the relationship, if any, among the anomalous
exponents $\zeta(p) , \sigma(p) , s(p)$.
We are now able to pose our problem in a quantitative way. 
Let us first consider the
very simple case $f_i = g_i$.
By subtracting 
equation (\ref{NS}) from  equation (\ref{NSW}) we obtain:
\be
\label{diff0}
\partial_t \phi_i + U_j \partial_j \phi_i =  - \partial_i \pi 
+\nu \Delta \phi_i 
\ee
where $\p = \W - \U$ and $\pi = (p-p_w)$. By
equation (\ref{diff0}) it immediately follow that the space 
average $E_\phi$ of $\p^2 = \int dx  \phi_i(x) \phi_i(x)$
 satisfies the equation:
\be
\partial_t E_\phi = - \epsilon_\phi 
\ee
where $\epsilon_\phi$ is the mean rate of dissipation of $E_\phi$. 
Thus, for long enough time, the field $\phi_i$ goes to zero and $W_i = U_i$,
identically.\\

In this paper we want to understand to which extent 
the statistical properties of (\ref{NSW}) are universal with respect
to the forcing mechanism. If a strongly universality holds, then we should have
 independently of the 
forcing mechanism 
\be
\sigma(p) = \zeta(p),\;\;\;  s(p) = \zeta(3p)\label{sca}
\ee
 even  when 
 $f_i$ and $g_i$ are uncorrelated or weakly correlated fields. Let us notice
that in the previous equality we have assumed that velocity flux 
possesses the same scaling properties of the velocity field, as always
verified in all numerical and experimental data.  
We do not know any rigorous argument against or in favor of universality
 for the scaling properties
of (\ref{NSW}). One may argue that if the forcing mechanisms $f_i$ and 
$g_i$ are weakly correlated --or  independent-- and
confined only to large scales than
the statistical properties of $\W$ may not be strongly influenced by the 
forcing itself. In
the latter case   the universality
of the scaling exponents of $\W$ should be  achieved by the same zero-modes 
mechanisms previously discussed for the passive scalar
with independent forcing. Whether this 
supposed universality can  be pushed until
the very extreme case of fully correlated systems, implying the equalities
 (\ref{sca})  is a matter of discussion. There is clearly a physical
relevant question here: we need to understand the importance of external 
forcing mechanism in the statistical properties of the advected field.
This question, of course, arises  not only for the passive
vector with pressure but also in any linear advection problem, i.e. passive
 scalars or passive magnetic fields. In the case
of the passive vector with pressure the question assumes also 
another interest because of the possibility to push the advected field to
follow the velocity field for some particular external forcing.

We consider an important point to investigate this problem in details and 
in the next section,
we present some numerical results showing that 
universality  holds for fluxes, i.e. $s(p) = \zeta(3p)$.
Passive structure functions seems to be less universal than fluxes
although strong
boundary effects (both ultraviolet and infrared) do not allow
a precise statement. 

\section{Numerical results}

Although a direct numerical  tests of the statistical properties
(\ref{NSW}) 
is possible for the  NS equations,  we will limit  in this paper to  
a direct  numerical investigation
 in the framework of
shell models for three dimensional turbulence.\\
Shell models for turbulent energy cascade
have proved to share many statistical properties
with both turbulent three dimensional velocity fields \cite{frisch,bjpv,sabra}
 and
with passive linearly advected quantities \cite{jpv}. Let us 
introduces  a set of wavenumber $k_n = \lambda^n k_0$ with 
$n = 0,\dots,N$ and the inter-shell ratio fixed to $\lambda=2$.
The shell-velocity variables $u_n(t)$ must be understood as 
the velocity fluctuation over a distances $r_n=k_n^{-1}$. 
We also introduce the linear-field shell variables $w_n(t)$
defined on the same sets of discrete wavenumber. It is possible
to write down a set of coupled  ODEs for the time evolution of
$u_n(t)$ and $w_n(t)$ which mimics the velocity and the
passive turbulent evolution (see below).  

We know that zero modes are at works in shell models
for linearly advected quantities, exactly like
in the true linear hydrodynamical problems. Indeed, it is 
possible
to prove analytically 
that  shell models for passive scalar advected by 
Gaussian and white-in-time velocity (shell) fields 
have intermittent corrections dominated by the null space of the linear
finite-dimensional advecting operator \cite{bw,bbw}. Recently,
it has also been shown that  shell models for passive
scalar advection, i.e. passive shells advected by 
shell fields arising from a shell model for the velocity,
have zero modes dominance for the scaling properties in the 
inertial shells, exactly as it is shown for the true passive
scalar \cite{cv00}.\\
We have therefore, exactly the analytical/phenomenological
framework useful to check  to which extent 
the scaling properties of the 
linear hydrodinamical problem are forcing independent. 

In the shell model framework the very meaning of pressure is absent.
The equivalent of our linear-hydrodynamical model (\ref{NSW}) will
become a shell field $w_n(t)$, linearly   advected by the shell model
velocity fields $u_n(t)$, conserving the energy $\sum_n |w_n|^2$, 
and such that when $f_n = g_n$ 
we have, for time large enough, $w_n(t)=u_n(t)$.

It is possible to write down such a coupled set of ODEs
for all shell models. Here we consider the case of Sabra model
\cite{sabra}. We obtain:
\be
(\frac{d}{dt}+\nu k_n^2)u_n = i(k_nu_{n+1}^*u_{n+2} + b k_{n-1}
u_{n+1}u_{n-1}^* + (1+b)k_{n-2}u_{n-2}u_{n-1}) +f_n
\label{shell_v}
\ee
\be
(\frac{d}{dt}+\nu k_n^2)w_n = i(k_n u_{n+1}^*w_{n+2} + b k_{n-1}
w_{n+1}u_{n-1}^* + k_{n-2}w_{n-2}u_{n-1}
+ b k_{n-2}w_{n-1}u_{n-2} ) +g_n
\label{shell_t}
\ee
where the non linear term of $u_n$ evolution and the linear advection
part of $w_n$ evolution have the only free parameter $b$.
Note that if we put $w_n = u_n$ in equation (\ref{shell_t}), then
equation (\ref{shell_t}) becomes equivalent to equation (\ref{shell_v}).
It is known that in order to have a realistic intermittent behavior
for $u_n$ one has to chose $-1<b<0$ \cite{sabra}.\\
 Equations (\ref{shell_v}) and
(\ref{shell_t}) preserve velocity energy, $E_u \equiv \sum_n |u_n|^2$,
and passive  energy,  $E_{w} \equiv 
\sum_n |w_n|^2$, in the limit of zero viscosity
and zero forcing. 

It is easy to  show that if $f_n=g_n$ in the above equations
 than for long enough time $w_n(t) \equiv u_n(t)$. 

The universality issue, as discussed in the previous section,
consists now in studying the scaling properties of 
$w_n$ at changing its forcing mechanism. \\
Let us
define  the scaling exponents 
for the velocity flux, $$\Pi_n= \Im [(k_nu_{n+2}u^*_{n+1}u_n^*)
			  +(1+b)k_{n-1}(u_{n+1}u_n^*u_{j-1}^*) ]$$
and  passive flux, $$Q_n=\Im [(k_n w_{n+2}u^*_{n+1}w_n^*)
			  +k_{n-1}(w_{n+1}u_n^*w_{n-1}^*) 
+k_{n-1} b (w_{n+1}w_n^*u_{n-1}^*)]$$
as:
\be
S^p_{\Pi}(n) \equiv \la|\Pi_n|^{p}\ra \sim k_n^{p-\zeta(3p)}
 \;\;\;\; T^p_{Q}(n) \equiv \la |Q_n|^{p} \ra \sim k_n^{p-s(p)}
\ee
where the equivalent of $4/5$ law for the two shell models 
gives $\zeta(3)=s(1)=1$, \cite{pbcfv}.
The  structure functions are defined as follows:
\be
S^p_u(n) = \la |u_n|^p \ra \sim k_n^{-\zeta(p)} \;\;\;\; T_w^p(n) =
 \la|w_n|^p\ra
\sim k_n^{-\sigma(p)}
\ee
In the following we present some numerical tests done with $N=25$ shells
$\nu_{u}=\nu_{w}= 5\cdot 10^{-7}$ and different kind of forcing mechanisms.\\
Forcing have been chosen such as to go from a weakly
correlated situation where $g_n(t)$ has some
large scale dependency from the $u_n(t)$ dynamics to a fully 
uncorrelated case with  $g_n(t)$ given by a random process.

In Fig. 1  we present the typical scaling laws one obtains for 
fluxes of both fields, for different
large scale passive-forcing.  In all 
simulations we have always taken the same velocity 
forcing concentrated
on the largest shell and constant: 
$f_n = (1+i) C_u \delta_{n,1}$, with $C_u=0.01$.  We have compared
statistical properties of the passive fields using  three
different choices for the passive forcing. 
 In case (A) we had a time-dependent forcing such as to 
impose a  constant
passive energy input on the first shell: 
 $$A: \;\;\; g_n(t) =  \delta_{n,1} (1+i)/w_1^*(t).$$
In case (B) we fixed the first passive shell
to have the same amplitude $|w_1|=const.$
 but leaving its phase to evolve according to its 
own dynamics.
$$
B: \;\;\; g_n(t) \rightarrow |w_1(t)| = 1. \;\; \forall \,t.
$$
In case (C) we took a forcing concentrated on the first shell,
with constant amplitude, $|g_1(t)|=G_1$ but random
independent phases
$$
C: \;\;\; g_n(t)  = \delta_{n,1} G_1 e^{i\theta(t)}
$$
where $\la \theta(t)\theta(t') \ra \propto \delta(t-t')$.\\
Let us notice that forcing of cases (A) and (B) have some (weak)
correlation with the advecting velocity field, while case (C) is
 independent of $u_n$. \\ 
As one can
see  all flux curves   
  superpose perfectly in the inertial range.
  Let us notice that the extremely small
errors
on the scaling of two fluxes allows us to support the 
statement $\Pi_n \sim Q_n$ with very high accuracy independently
on the forcing mechanism. 
 The  scaling of passive structure functions 
suffers of larger error bars, due to a
less smooth matching between inertial and infra-red properties
\cite{note1}. The qualitative trend is, anyhow, toward a more
intermittent behavior of the passive fields with respect to
the velocity field (see Fig. 2). In the latter case
we would have an indication that passive structure
functions are much more sensible to the boundary conditions
than the fluxes moments, implying also a weaker degree of universality
with respect to the forcing mechanism.\\
In Table 1 we quantify our finding in all three forcing cases (A-C) 
showing the best fits and their errors for all scaling exponents
of both fluxes and structure functions.\\
In Fig. 3, we show two typical time evolution for the total
energies of velocity, $E_u(t)$, and passive, $E_{w}(t)$
for the case with two different -but correlated- forcings, $f_n \neq g_n$.
Although some correlation between $E_u(t)$ and $E_{w}(t)$
is observed, we are very far from the trivial exact correlated
case one would have obtained choosing  $f_n \equiv  g_n$. The non
trivial correlations between the two energy can be seen in the
fig. 4 where we plot, for a typical
trajectory, $E_u$ versus $E_{w}$.\\
In order to test the robustness of previous results we repeated
the numerical experiments with a different value of the
free parameter $b$ in the sabra shell model equations (\ref{shell_v}).
 It is known
that at changing $b$ the intermittency of the velocity field changes.\\
In Fig. 5 we plot the $\zeta(3p)$ and $s(p)$ curves for both
values of $b=-0.4$ and $b=-0.6$. The agreement for each $b$ values
between the two fluxes is again perfect despite the fact that
at varying  $b$ we have different degrees of intermittency.

One could question why there is such clear evidence that 
$s(p) = \zeta(3p)$,
i.e. the scaling properties
of two fluxes are identical, while passive structure functions 
seem to be more intermittent than the corresponding velocity 
structure functions. One possible explanation goes as follows. 
By exploiting the equation of motion one may always derives
homogeneous constraints for moments of 
quantities like  $F_{\n}= u_n w_{n'} w_{n''}$
in the inertial range, i.e. where forcing is not directly acting.
  In particular by writing the stationary condition
for quantities  like $ \la F_{\n_1}F_{\n_2}\cdots F_{n_{p-1}} w_k w_{k'} \ra$
one obtains an homogeneous constraints involving only $F_{\n}$ observable:
\be
\frac{d}{dt} \la F_{\n_1}F_{\n_2}\cdots F_{\n_{p-1}} w_k w_{k'} \ra = 
O_{\n_1\n_2\cdots \n_{p-1},k,k'}^{\m_1\m_2\cdots \m_{p}} \la 
F_{\m_1}F_{\m_2}\cdots F_{\m_{p}}\ra =0
\label{zero}
\ee
where the operator 
$O_{\n_1\n_2\cdots \n_{p-1},k,k'}^{\m_1\m_2\cdots \m_{p}}$
is given from the equation of motion and it is
independent of the correlation between $u_n$ and $w_n$ and of the 
chosen forcings $g_n$ and $f_n$.  
 On the other hands no homogeneous closed constraints can ever be found
for moments of quantities involving only simultaneous correlations
of passive fields, $w_n$.
In order to have homogeneous expressions for obervable 
made of only  passive fields
one has to give (or to guess) an explicit form for
the correlation between $w_n$ and $u_n$. Thus one may think to obtain
an equation similar to (\ref{zero}) but now with 
the inertial operator  explicitly dependent on the given
correlation between the two fields. In the latter case
the dependency of the passive field on the statistical
properties of the velocity field and of the correlation
with the external forcing may lead to non-universal scaling
properties.\\
On the other hand, one can also argue that the observed   lack 
 of universality
is not Reynolds independent  and that at Reynolds large enough
some strong independence from the forcing mechanism (zero-modes
dominance) would be recovered
also for passive structure functions. 
The latter  scenario is what happens  in true
passive scalars with a correlation between forcing and velocity
field (see the case of a passive scalar with shear \cite{lanotte}) where the
existence of
 sub-leading
non-homogeneous terms induced by the forcing mechanism may spoil the 
scaling behavior of the zero-modes at finite Reynolds numbers. 

\section{An approximate computation of the scaling exponents}

The results so far discussed, make us 
rather confident that the two fluxes  of equation 
(\ref{NSW})  and of  Navier Stokes eqs (\ref{NS})
have the same statistical fluctuations. 

We want now to understand whether it is possible to use the above results
 to obtain useful informations
on the scaling exponents $\zeta(p)$.

The main idea, discussed in this section, is the following. We assume that
the velocity field can be described by the (unknown) 
multifractal probability distribution
$P(\delta_rU)$. We want to compute the probability distribution of the 
passive vector, $P(\delta_rW)$.  The equation 
(\ref{NSW}) will induce a functional
relation between the two probability distributions. By requiring that
the scaling
properties of $\W$ flux  are the same of $\U$ flux,
 we introduce an (infinite) set of
equations for the probability
  $P(\delta_rU)$ whose solutions will fix its scaling.
In order to simplify the above procedure, the approach we want to follow is
based on a suitable set of 
approximations and assumptions
whose validity we are not able to prove
rigorously, only {\it a posteriori} we will be able to judge the goodness
of our  calculation.\\
Let us suppose that the field $\U$ in equation (\ref{NSW})
is characterized by a multifractal spectrum $D(h)$. As it is well known, 
in such a case the anomalous
exponents of $\U$ are given by the expression 
$\zeta(p) = \inf_h ( ph+3-D(h))$ and the probability
distribution of $\delta_r U$ is given by \cite{bbpvv}:
\be
\label{PDFU}
P(\delta_rU) \propto \int d\mu(h) r^{3-D(h)} \exp\left[-\frac{(\delta_rU)^2}{2U_0r^{2h}}\right] 
\ee
where $U_0$ represents the variance of the large scale velocity field, 
which is supposed to be Gaussian.
Equation (\ref{PDFU}) tells us that for each value of $r$ we may 
consider $\delta_rU$ as the superposition
of Gaussian field with variance $U_0r^{2h}$ and probability $r^{3-D(h)}$ 
associated to each
value of $h$. We can then consider to solve equation (\ref{NSW}) in the 
limit where $\U$ is a "weighted" 
superposition of Gaussian random fields (as expressed by (\ref{PDFU})).\\

The equivalence between the statistical properties of the Navier-Stokes
field and of the linear field implies that
both fluxes have the same scaling:
\be
\la |(\delta_r U) (\delta_r W)^2|^p\ra \sim \la |\delta_r U|^{3p} \ra
\label{fluxes}
\ee
We want to exploit the above identity in order
to derive a constraint for the $D(h)$ spectrum.\\
We can rewrite  the term $\la(\delta_rU(\delta_rW)^2)^p\ra$
as follows:
\be
\la (\delta_rU(\delta_rW)^2)^p\ra
 \sim \int dh  d(\delta_r U) 
r^{3-D(h)} P(\delta_r U,h) (\delta_r U)^p
\la (\delta_rW)^{2p}|\delta_r U, h\ra
\label{cond}
\ee
where with $ \la (\delta_rW)^{2p}|\delta_r U,h \ra$ we mean 
the average of the linear field conditioned to the advecting
velocity field and $P(\delta_r U,h) \equiv 
\exp\left[-\frac{(\delta_rU)^2}{2U_0r^{2h}}\right]$. 
Up to eqn. (\ref{cond}) we did not used any
approximation.
\noindent
The computation of $ \la (\delta_rW)^{2p}|\delta_r U,h\ra$
 is the most difficult one and we shall introduce several
approximations in order to make it feasible. 
In particular we want to compute
$ \la (\delta_rW)^{2p}|\delta_r U,h\ra$ using a surrogate $\delta$-correlated
and Gaussian velocity field. In other words 
 will make the first approximation that in order to compute the conditional
average we may assume that 
the multifractal velocity field, as defined by equation (\ref{PDFU}), is the
"weighted" 
superposition of independent Gaussian random fields  
$\delta$ -correlated  in time.\\ 

In order to compute left hand side of equation (\ref{cond}),
we  solve
equation (\ref{NSW}) for fixed exponent $h$ of the 
random field $\U$ and then 
average the
results over $h$ with probability $r^{3-D(h)}$. In doing such an average we
 notice 
that in order to mimic the advection by a true Navier-Stokes
field (not $\delta$-correlated in time) with scaling  $\delta_r U \sim r^h$
we need a $\delta$-correlated surrogate with 
scaling $\delta_r U_{s} \sim r^{\frac{1+h}{2}}$,
because of dimensions carried by the delta-functions. As a consequence
we are going to be interested only in exponents
for the surrogate $U_{s}$ field in the range
 $H = \frac{1+h}{2} = [1/2,1]$, which correspond
to the range $h:[0,1]$ for the true turbulent field. Let us notice
that a K41 field has a correspondent $\delta$-correlated field
scaling as  $\delta_r U_{s} \sim r^{2/3}$.\\

We now want to exploit our $\delta$-correlated
ansatz  by writing:
\be
\la (\delta_rW)^{2p}|\delta_r U, h\ra \sim r^{ p(1-h) +\rho_{2p}(1+h)}
\label{cond2}
\ee
where we have introduced the anomaly $\rho_{2p}(2H)$
of the linear field $\W$ advected by a $\delta$-correlated velocity
field $\delta_r U_{s} \sim r^H$, namely:
\be
\label{H}
\la (\delta_rW)^{2p}) \ra _H \sim r^{ p(2-2H) +\rho_{2p}(2H)}
\ee
In equation (\ref{H})
 we have taken into account the dimensional 
consistency relation $2H=1+h$. Notice that for the $\delta$-correlated
case one can prove that in the isotropic sector we have $\rho_2(2H)=0$,
i.e. the second order correlation function has not anomalous 
scaling.\\
Our definition gives:
\be
\la \delta_rU(\delta_rW)^2\ra_H \sim r^H r^{1-H} =  r
\ee
i.e, our ansatz on the function $ \la (\delta_rW)^{2p}|\delta_r U,h\ra$ 
 implies that we are averaging
over all possible singularity $H$ with the constraint 
$\la \delta_rU (\delta_rW)^2\ra_H\sim r$.\\
We can now use the relation of statistical identity between
fluxes,  $\zeta(3p) = s(p)$, in order to obtain an 
equation for
$D(h)$. We have:
\be
\int d\mu(h) r^{ph+p(1-h)-\rho_{2p}(1+h) +3 -D(h)} \sim r^{\zeta(3p)}
\ee
which gives:
\be
\label{DH}
\inf_h\left[ p -\rho_{2p}(1+h) + 3-D(h)\right] = \inf_h \left[ 3ph + 3 -D(h)\right]
\ee
If we are able to compute the anomalous 
correction $\rho_{2p}$ (of the $\delta$-correlated problem), 
equation (\ref{DH}) becomes
a functional equation for $D(h)$ whose solution 
gives us the anomalous exponents $\zeta(p)$
within the set of approximation discussed above.
 Let us notice that (\ref{DH}) is consistent with
the known result $\tau(1) = s(1) = 0$.
Let us also remark that as soon as we introduce any smoothing
in the time dependency of $\U$ 
the equation (\ref{DH}) may be no longer valid and we should reconsider
the averaging procedure in a more suitable way.\\
The computation of the anomalous 
exponents $\rho_{2p}$ is a feasible but difficult 
task for equation (\ref{NSW}) because 
of the condition $\partial_i W_i = 0$. In principle,
the computation can be done perturbatively following the
analog of the passive scalar case \cite{gk95}. 
Only the exponents in different anisotropic sectors
for the second order correlation function have been, up to now,
computed.  However, in order to understand the 
quality of the approximations so
far introduced to derive 
equation (\ref{DH}), at least for three dimensional
isotropic and homogeneous turbulence, we can use the numerical
 values of $\rho_4(1+h)$ and $\rho_6(1+h)$ as recently 
computed numerically for
the passive scalar model \cite{massimo}, hoping that the introduction
of the pressure term does not change too much the scaling
exponents \cite{note2}.

In order to find a solution of (\ref{DH}),
 we simply assume that 
$D(h)$ can be parameterized by using the 
expression $D(h) = d_0 (1-x+xln(x))$
where $x = (h-h_0)/(d_0log(\beta^{-1}))$ which corresponds to 
a log-poison distribution for the multifractal model. We use a log-poison
formula because we know that it parametrize particularly well the experimental
data \cite{exp,exp1}. 
Because $d_0$ is fixed by
the condition $\zeta(3) =1$,
 we are left with the problem to find $\beta$ and $h_0$ by using
equation (\ref{DH}) for $p=2$ and $p=3$. 
 Solving equation (\ref{DH}) for $\beta$ and $h_0$ gives
$\beta = 0.781$ and $h_0 = 0.14$. In fig. 6
we show the estimated values of $\zeta(p)$
for $p=1,..12$ against the experimental
 findings for homogeneous and isotropic turbulence. 
We note that the
 estimate based on
 (\ref{DH}) is rather accurate with an error not greater than $5 \%$.
We argue that such a small error shows that our approach, within 
the limitation and the approximations
previously discussed, looks promising as an 
useful tool to compute the anomalous exponents.

In order to test the sensitivity of (\ref{DH}) to the 
values of $\rho_4(1+h)$ and $\rho_6(1+h)$, it is interesting
to use the anomalous exponents 
$\rho_4$ and $\rho_6$ given by Kraichnan' formula for $D=3$.
In this case the solution 
of (\ref{DH}) gives $\beta = 0.813$ and $h_0 = 0.21$ and 
the corresponding values of $\zeta(p)$ are also plotted in fig. 6.
Although there are not big differences
between   exponents obtained using the  Kraichnan formula
and those obtained by using the correct numerical results, the agreement
with the experimental data
is definitely better in the latter case. 
\section{Conclusions}
In this paper we discussed several properties
 characterizing the statistical behavior
of a divergence-less vector field passively 
advected by an homogeneous and isotropic 
turbulent field. 
\noindent
We can summarize our findings in the following way.
\noindent
(i) We present some theoretical 
arguments which supports the statement that the flux of passive vector field
should have the same statistical properties
 of the flux of an homogeneous and isotropic turbulent field.
(ii) We generalize the concept of divergence-less
 vector field to shell models and by using detailed numerical 
simulations we provide evidence that the 
anomalous scaling exponents of  the passive vector flux and of the 
non linear shell model flux are the same with very high accuracy.
(iii) We propose a self consistent approach to 
compute the anomalous exponents in homogeneous and isotropic turbulence by
using the properties previously discussed. 
In particular, we have been able to define a functional equation for
$D(h)$ within a suitable set of approximations.
(iv) By assuming that the pressure term does not
 change dramatically the numerical values of the anomalous exponents
for the divergence-less vector field advected
 by a Gaussian isotropic $\delta$-correlated random field, 
we have been able
to find an approximate solution of the functional 
equation for $D(h)$ which compares rather well to known
experimental data.

Before closing we want to discuss a generalization 
of the model
(\ref{NSW}) which we consider interesting for further studies. Let us 
consider the 
following set of equations:
\be
\label{NSUL}
\partial_t  U_i +  U_j  \nabla_j U_i + \lambda  W_j \nabla_j U_i = 
-\nabla_i p_u + \nu \Delta  U_i +  f_i
\ee
\be
\label{NSWL}
\partial_t  W_i +  U_j  \nabla_j  W_i + \lambda  W_j  \nabla_j  W_i =
 -\nabla_i p_w + \nu \Delta W_i +  g_i
\ee
where we assumed that $\partial_i U_i = \partial_i W_i = 0$.
It is interesting to notice that the  vector field $ \Z_{\lambda} = \U + \lambda \W$ 
satisfies the Navier Stokes equations and that  equations (\ref{NSUL})
 and (\ref{NSWL}) corresponds to
 (\ref{NS}) and (\ref{NSW}) for $\lambda = 0$. 

Finally we want to remark that
 equation (\ref{NSW}) is an useful tool to understand the role played 
by coherent structure on the anomalous scaling 
in both two dimensional and three dimensional turbulence. 
All our findings suggest that a systematic
 study of equation (\ref{NSW}) looks extremely promising
in order to derive a new approach in understanding intermittency 
and estimating  anomalous scaling in Navier-Stokes 
equations.\\
We acknowledge useful discussion with I. Arad, A. Celani and  M. Vergassola.
 We thank M. Vergassola
and A. Mazzino who provided us the numerical data for the
4th and 6th order anomaly in the Kraichnan passive
scalar problem. This work has been partially supported by
the EU under the Grant 
No. HPRN-CT  2000-00162 ``Non Ideal Turbulence'' and by the INFM  (Iniziativa
di Calcolo Parallelo).

\newpage
\begin{table}[htb]
\begin{tabular}{||c|c|c|c|c|c|c|c||}
    p  & $k_n^{-p/3} S_{\Pi}^{p/3}(n) $  & $k_n^{-p/3} T_{Q}^{p/3} (x) $ (A) &  $k_n^{-p/3} T_{Q}^{p/3}(n)$ (B)&
  $k_n^{-p/3} T_{Q}^{p/3}(n)$ (C) & 
$T_w^p(n)$   (A) &  $T_w^p(n)$  (B) &  $T_w^p(n)$  (C) \\
\tableline
 2  & 0.712 (3)  & 0.711 (3) & 0.711 (2) &  0.710 (2) & 0.67 (3)  &  0.67 (3) & 0.66 (3) \\
 4  & 1.263 (6) &   1.264 (7) &  1.264 (6)& 1.266 (3)  & 1.20 (6) & 1.28 (4)&  1.13 (7)\\
 6  & 1.745 (8)  &  1.741 (7)   & 1.74 (1) &  1.745 (5)  & 1.6 (1) & 1.67 (8) &  1.4 (1) \\
 8  & 2.18 (2) &  2.18 (2)   & 2.19 (2)  &  2.18 (2)  &  2.0 (2) &  2.1 (1)&  1.6 (2) \\
10   & 2.60 (2) &  2.58 (3)   & 2.61 (2) &  2.57 (4)  &  2.3 (2) & 2.6 (2)  & 1.9 (3)  \\
\end{tabular}
\caption{Scaling exponents of both flux and structure function of
the passive field and of flux of velocity field at changing the order
of the moment $p=2,4,6,8,10$ and for different large scale forcing 
cases (A-C). Notice  the high precision in the 
agreement
between the passive flux at changing the large-scale forcing (columns 1--4). 
Passive
structure functions  (columns 5--7) are less accurate due to the presence
of a strong bottleneck at small scales. Errors are given by the 
numbers in brackets and refers to the last digit. Errors
are estimated from the 
fluctuations of the local logarithmic 
derivatives in the shell range $ n = 3,\dots,15$.}
\end{table}
\newpage
\begin{figure}
\hskip -.4cm\epsfig{file=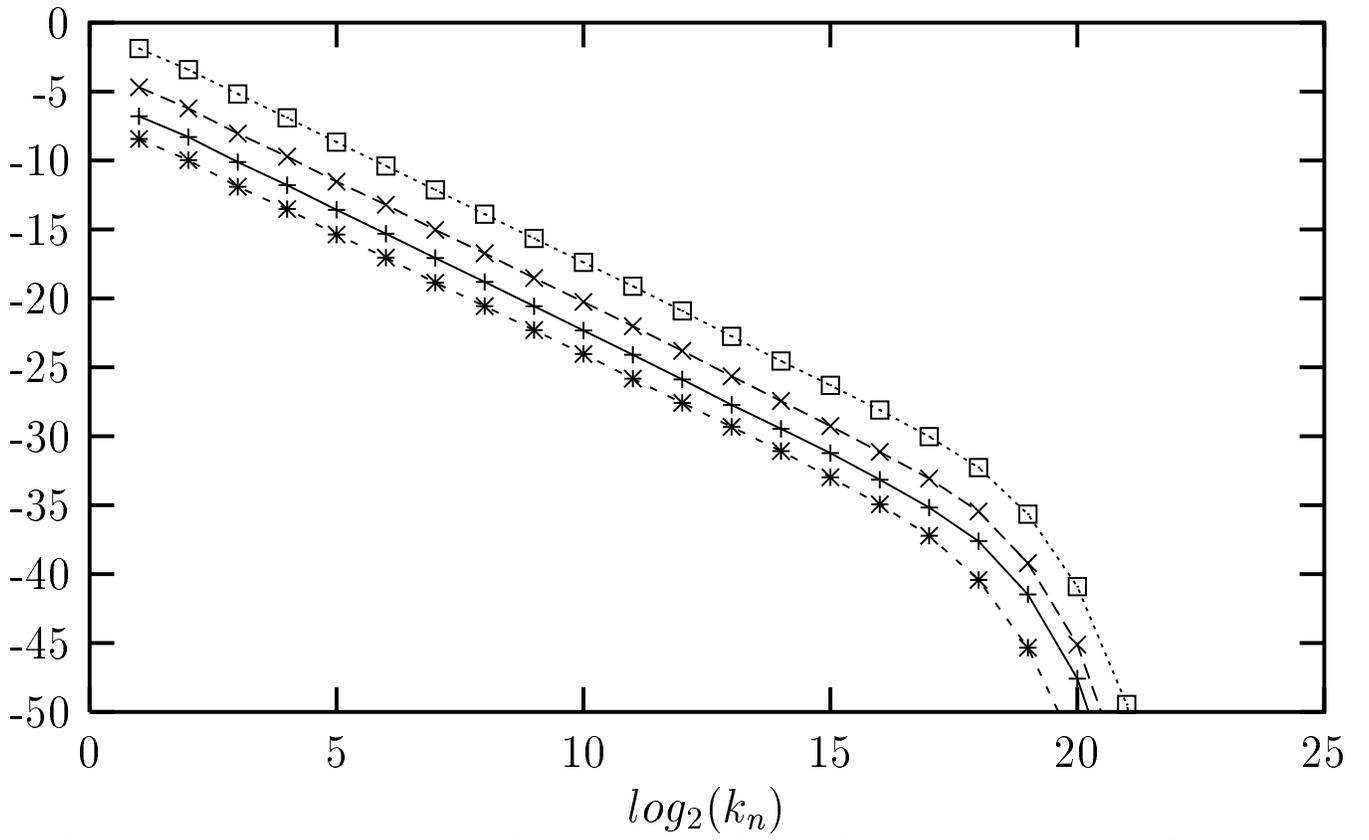,width=\hsize}
\caption{Log-log plot 
of velocity and passive fluxes and structure functions for
second order moment, $p=2$ versus the scale $k_n$. Curves represent from above:
 ($\Box$) velocity flux $S_{\Pi}^2(n)$. Passive flux, $T_{Q}^2(n)$,
for   forcing case (A) ($\times)$;   forcing case (B) ($+$); forcing case (C)
($\ast$).  Fluxes are 
always multiplied by the normalising factor $k_n^{-2}$. Curves
have been shifted along the y-axis for the sake of clarity}
\end{figure}

\begin{figure}
\hskip -.4cm\epsfig{file=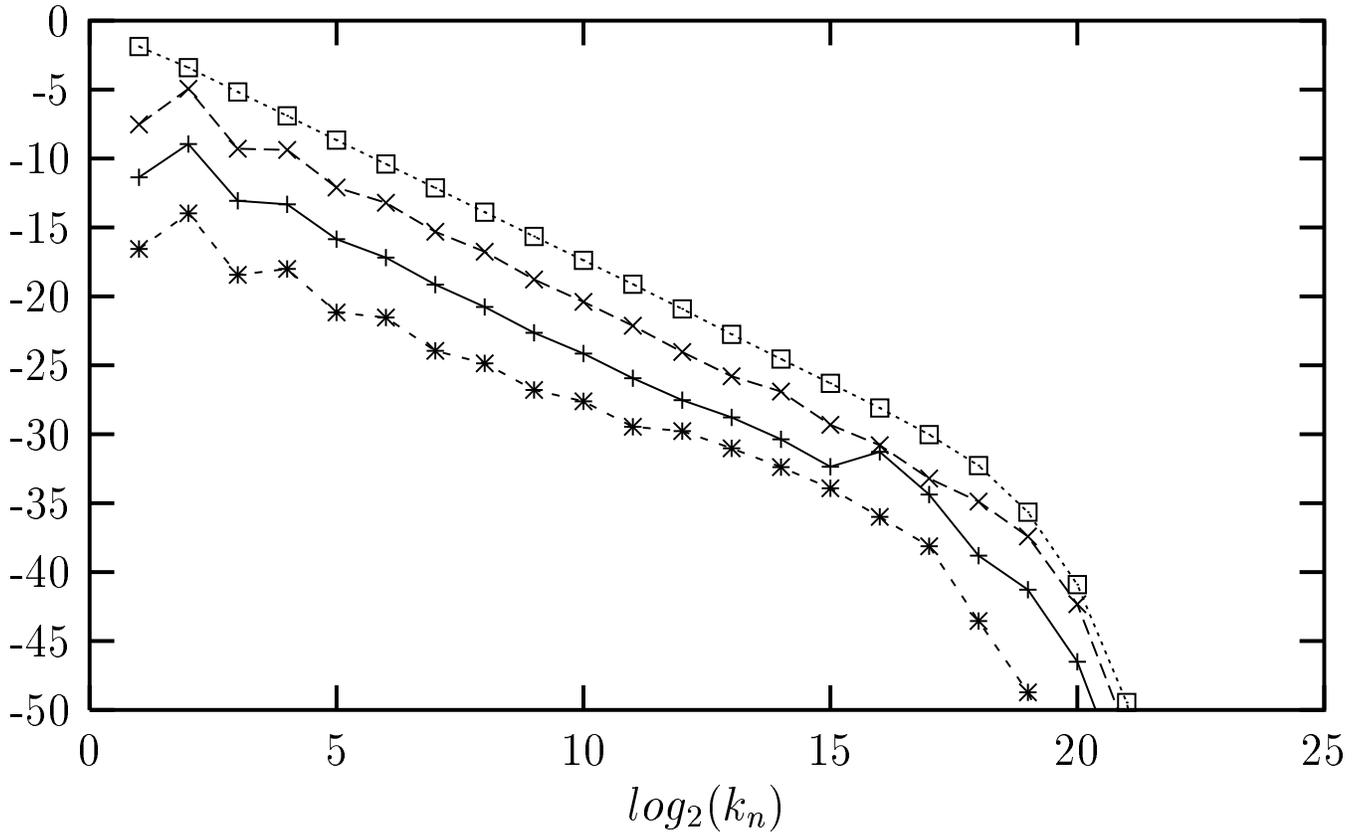,width=\hsize}
\caption{Log-log plot
of second order velocity  flux, $p=2$, and sixth order 
passive structure functions,  versus the scale $k_n$. 
Curves represent from above:
 ($\Box$) velocity flux $S_{\Pi}^2(n)$; ($\times$)
  passive  structure function $T_{w}^6(n)$ forcing case (A); ($+$)
  passive  structure function  $T_{w}^6(n)$ forcing case (B); ($\ast$)
  passive  structure function $T_{w}^6(n)$ forcing case (C). Notice the
strong ultraviolet and infrared effects in the passive structure functions.
Notice also that  for the fully independent forcing, case (C),
the passive structure functions shows a larger intermittent
slope at small scales. }
\end{figure}

\begin{figure}
\hskip -.4cm\epsfig{file=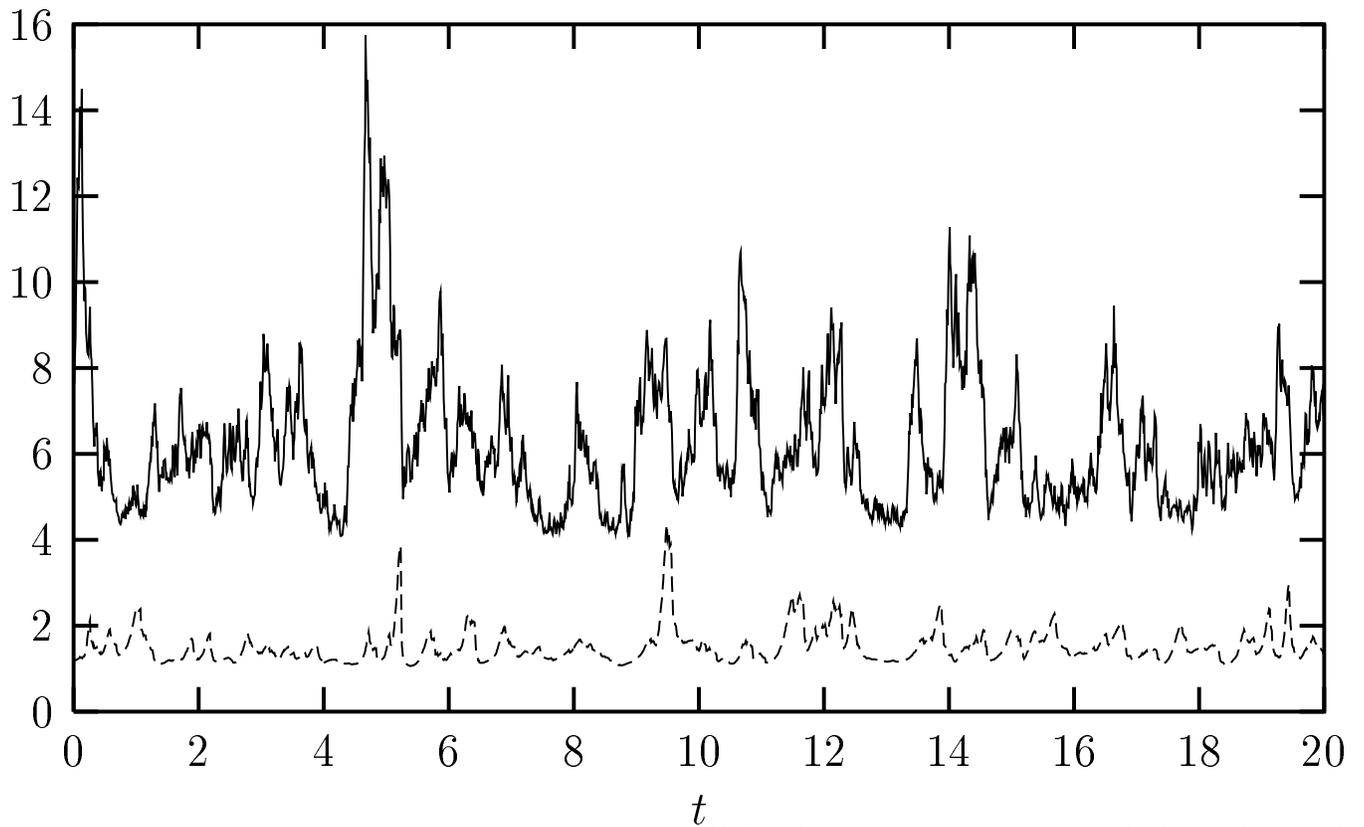,width=\hsize}
\caption{Typical time evolution of the total velocity energy $E_u(t)$ (solid)
and total passive energy $E_w(t)$ (dashed) (case A). Curves
have been shifted along the y-axis for the sake of clarity}
\end{figure}
\newpage

\begin{figure}
\hskip -.4cm\epsfig{file=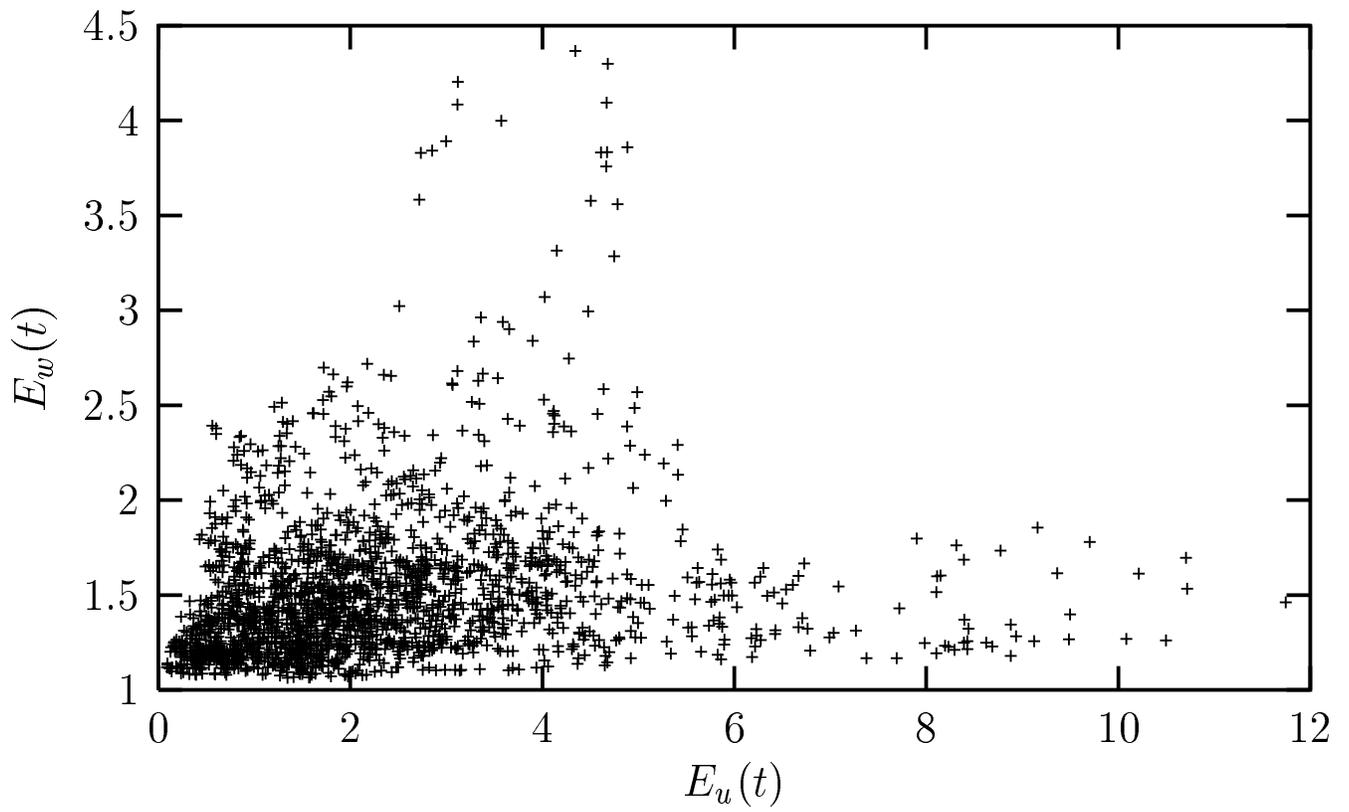,width=\hsize}
\caption{Scatter plot of  $E_u(t)$ versus  $E_w(t)$ with forcing
of case (A).
Notice that the two fields have not perfect correlation.}
\end{figure}
\newpage

\begin{figure}
\hskip -.4cm\epsfig{file=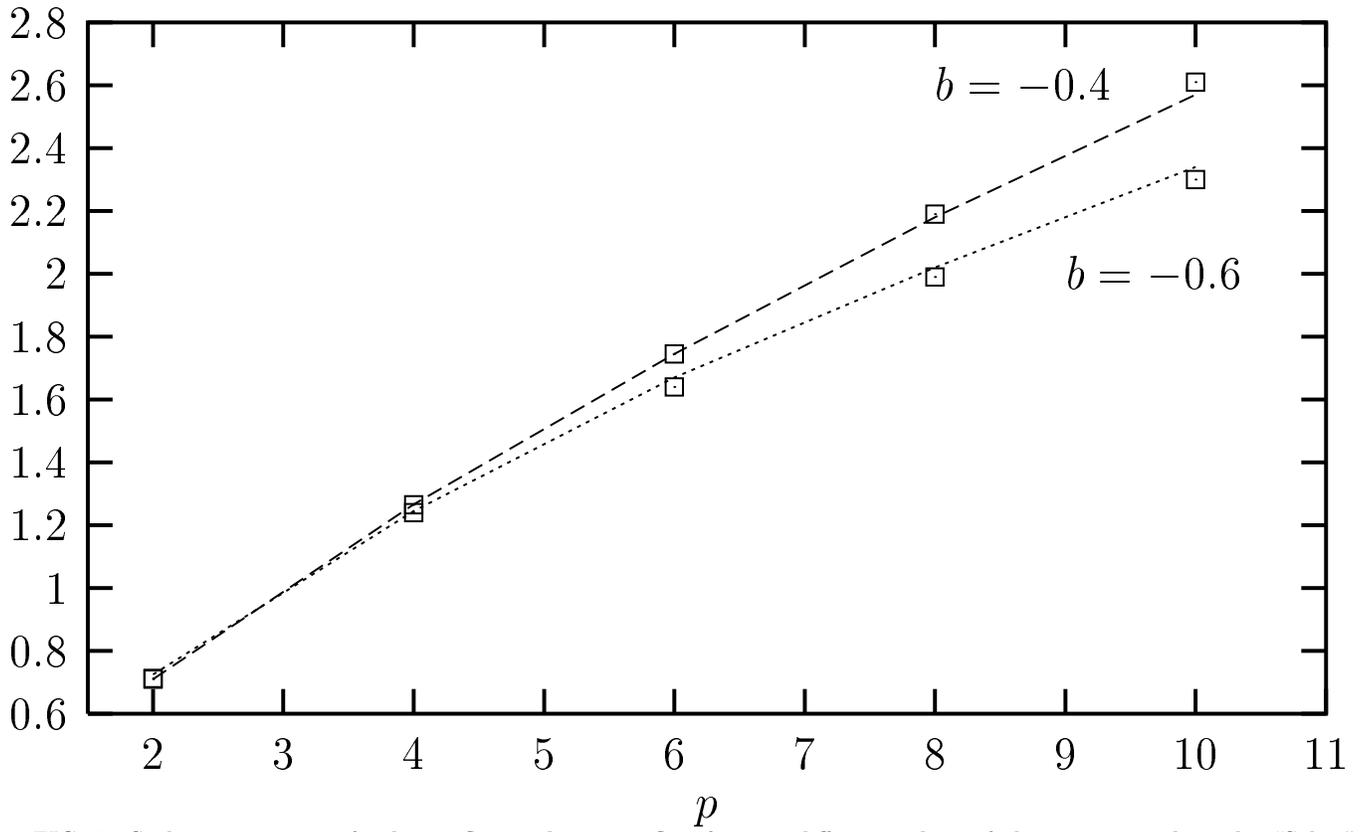,width=\hsize}
\caption{Scaling exponents of velocity flux and passive flux %
for two different values of the parameter  $b$ in the ``Sabra'' 
shell model equations (\ref{shell_v}). $\Box$ corresponds
to the velocity flux exponents, $\zeta(p/3)$,  and the dashed curves to the
passive flux exponents, $\sigma(p)$. Above: the case with $b=-0.4$, below
the case with $b=-0.6$. Notice that despite the remarkable change
in the intermittent properties the fluxes follow eachother perfectly.}
\end{figure}

\begin{figure}
\hskip -.4cm\epsfig{file=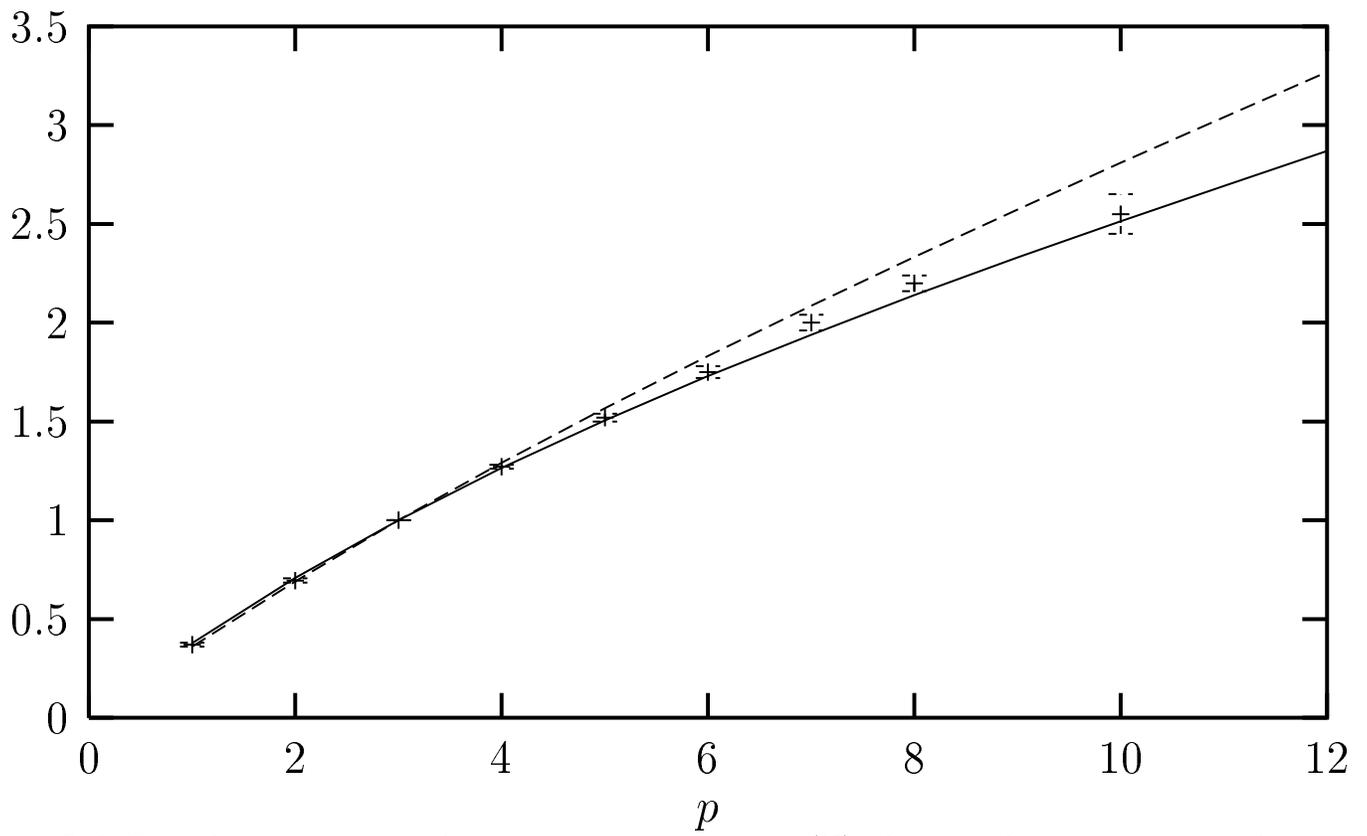,width=\hsize}
\caption{Comparison between the scaling exponents computed
 by  eqn. (\ref{DH}) using: numerical results for the passive
 scalar [17] (solid curve); the Kraichan formula for the passive scalar [1] (dashed curve).
Experimental data ($+$)  are given with their error bars.}
\end{figure}


\begin{thebibliography}{99}
\bibitem{note1} Such a different behavior of matching between structure
functions and fluxes is common to all shell models. It is meanly
due to the fact that the ``true'' scaling observable here are
only fluxes.
\bibitem{note2} Let us stress again that this is exactly true for the
second order correlation function.
\end{thebibliography}
\end{document}